\documentclass[aps,prxquantum,reprint,superscriptaddress]{revtex4-1}
\usepackage[utf8]{inputenc}

\usepackage{graphicx}  
\usepackage{xcolor}
\usepackage{bm}        
\usepackage{amssymb} 
\usepackage{amsmath}
\usepackage{amsthm}
\usepackage{amsfonts}
\usepackage{caption}
\usepackage{ragged2e}
\DeclareCaptionJustification{justified}{\justifying}
\usepackage[demo,abs]{overpic}

\newcommand{\ket}[1]{\left|#1\right\rangle}
\newcommand{\bra}[1]{\left\langle#1\right|}

\begin{document}
\title{Adiabatic quantum imaginary time evolution}

\author{Kasra Hejazi}
\affiliation{Division of Chemistry and Chemical Engineering, California Institute of Technology, Pasadena, California 91125, USA}

\author{Mario Motta}
\affiliation{IBM Quantum, Almaden Research Center, San Jose, California 95120, USA}

\author{Garnet Kin-Lic Chan}
\affiliation{Division of Chemistry and Chemical Engineering, California Institute of Technology, Pasadena, California 91125, USA}

\begin{abstract}
We introduce an adiabatic state preparation protocol which   implements quantum imaginary time evolution under the Hamiltonian of the system. 
Unlike the original quantum imaginary time evolution algorithm, adiabatic quantum imaginary time evolution does not require quantum state tomography during its runtime, and unlike standard adiabatic state preparation, the final Hamiltonian is not the system Hamiltonian. Instead, the algorithm obtains the adiabatic Hamiltonian by integrating a classical differential equation that ensures that one  follows the imaginary time evolution state trajectory. We introduce some heuristics that allow this protocol to be implemented on quantum architectures with limited resources. We explore the performance of this algorithm via classical simulations in a one-dimensional spin model and highlight essential features that determine its cost, performance, and implementability for longer times,
and compare to the original quantum imaginary time evolution for ground-state preparation.
More generally,  our algorithm expands the range of states accessible to adiabatic state preparation methods beyond those that are expressed as ground-states of simple explicit Hamiltonians.
\end{abstract}

\maketitle

\section{Introduction}

A central step in the quantum simulation of physical systems is to prepare a relevant initial state. Taking  ground-state simulation as an example, we typically wish to prepare a state with sufficient overlap with the desired ground-state $\ket{\Psi_0}$ of a Hamiltonian $H$. In the context of near-term quantum algorithms \cite{preskill2018quantum}, which minimize both qubits/ancillae and gate resources, many protocols for such ground-state preparation have been proposed. Examples include variational ansatz preparation~\cite{bauer2020quantum,peruzzo2014variational,farhi2014quantum,mcclean2016theory,schon2007sequential,romero2018strategies}, adiabatic state preparation (ASP) \cite{albash2018adiabatic,babbush2014adiabatic,veis2014adiabatic}, and variational~\cite{mcardle2019variational} and quantum imaginary time evolution (QITE) \cite{motta2020determining,sun2021quantum,yeter2020practical,gomes2020efficient,nishi2021implementation,kamakari2022digital,yeter2022quantum}, the latter being the subject of this work.

Because ground-state preparation is in general formally hard, all these methods rely on some assumptions. For example, ASP starts from an initial Hamiltonian $H(0)$, whose ground-state is simple to prepare, and defines an adiabatic path  $H(s), 0\leq s \leq  1$ with $H(1)\equiv H$, the desired Hamiltonian \cite{farhi2000quantum}. To prepare the ground-state to sufficient accuracy, $H(s)$ must change slowly; an estimate of the adiabatic runtime is \cite{messiah2014quantum,albash2018adiabatic}  $T \sim \max_{s,j} \frac{\left| \left\langle \Psi_j(s)\right| dH/ds \left| \Psi_0(s) \right\rangle \right|}{\Delta^2}$, where $\left| \Psi_0(s) \right\rangle$, $|\Psi_j(s)\rangle$, with $j\geq1$, denote the instantaneous ground and excited states of $H(s)$ and $\Delta(s)$ is the energy gap between the ground-state and the first excited state. 
For ASP to be efficient, $H(s)$ must be chosen such that $\min_s \Delta(s)$ is not too small, e.g. at worst $1/\mathrm{poly}(L)$ in system size $L$, for a polynomial cost algorithm.

The QITE algorithm~\cite{motta2020determining} on the other hand,  applies  $e^{-H\tau}$ ($\tau >0$) to boost the overlap of a candidate state 
with the ground-state of $H$; for this work, we consider Hamiltonians that are sums of local terms $H = \sum_\alpha h_\alpha$, where $h_\alpha$ is geometrically local 
(i.e. each term $h_\alpha$ acts on a constant number of adjacent qubits regardless of system size).
Ref.~\onlinecite{motta2020determining} introduced a near-term quantum algorithm to obtain the states 
\begin{align}
|\Psi(\tau)\rangle = e^{-H \tau}|\Psi(0)\rangle \ / \ \left\lVert e^{-H\tau} \ket{ \Psi(0) } \right\lVert, \label{eq:imag_time_evolved_psi}
\end{align}
without employing any ancillae or postselection. The method is efficient if $|\Psi (\tau)\rangle$ has finite correlation volume $C$ for all earlier imaginary times, in which case $|\Psi (\tau)\rangle$ can be prepared
by implementing a series of local unitaries acting on $O(C)$ qubits on the candidate state. By using this technique which reproduces the imaginary time trajectory, one can also use QITE as a subroutine in other ground-state algorithms, as well as to prepare non-ground-states and thermal (Gibbs) states, for example by reintroducing ancillae~\cite{kamakari2022digital}, or by sampling~\cite{motta2020determining}.

However, to find the unitaries in QITE
one needs to perform
tomography~\cite{motta2020determining} of the reduced density matrices of $|\Psi(\tau)\rangle$ over regions of volume $C$. Although the measurement and processing cost is polynomial in system size, it can still be prohibitive for large $C$. Despite various improvements in the QITE idea in terms of the algorithm and implementation, this remains a practical drawback~\cite{sun2021quantum}. (We briefly note also some other near-term imaginary time evolution algorithms, such as the variational ansatz-based quantum imaginary time evolution, introduced in Ref.~\cite{mcardle2019variational}, 
which reproduces the imaginary time evolution trajectory in the limit of an infinitely flexible variational ansatz, as well as the probabilistic imaginary time evolution algorithm (PITE)~\cite{kosugi2022imaginary}, whose probability of success decreases exponentially with evolution time).

Here, we introduce an alternative near-term, ancilla-free, quantum method that generates the imaginary time evolution of a quantum state without any tomography. It thus eliminates one of the resource bottlenecks of the original QITE. The idea is to consider the imaginary time trajectory $\Psi(\tau)$ as generated by an adiabatic process  under a particular Hamiltonian $\tilde{H}(\tau)$. This adiabatic Hamiltonian is approximated by the solution of an auxiliary dynamical equation that can be solved for entirely classically, i.e. without any feedback from the quantum simulation. 
Although propagation under $\tilde{H}$ reproduces imaginary time evolution when performed adiabatically (i.e., one stays in the ground-state of $\tilde{H}(\tau)$), this is different to the usual ASP, because  $\tilde{H}$ does not  approach $H$ at the end of the path, even though it shares the same final ground-state. We thus refer to this algorithm as adiabatic quantum imaginary time evolution, or A-QITE.

Like QITE, the application of  A-QITE  for sufficiently long imaginary time formally prepares the ground-state. But also like QITE,  the imaginary time trajectory generated by A-QITE has other applications. The ability to reproduce non-unitary evolution enables applications to Lindblad simulation~\cite{kamakari2022digital}, while for finite imaginary time,  
it can naturally be applied to prepare thermal states (either in a purified formulation~\cite{kamakari2022digital} or via sampling~\cite{motta2020determining,sun2021quantum}) or be used in a subspace expansion method for excited states~\cite{motta2020determining}. Generally speaking, the A-QITE algorithm turns the preparation of any state that can be reached by the differential evolution of an initially known ground-state into an adiabatic state preparation process. In this work, however, we primarily focus on the generation of the imaginary time trajectory itself, and its long-time limit relevant to physical ground-state preparation.

We examine the feasibility and performance of A-QITE for 
the illustrative case of 
the {Ising-like} Heisenberg XXZ model in a transverse field. 
There we study the behaviour of the instantaneous gap and norm of $\tilde{H}$ as a function of imaginary time (as these determine the cost of integrating the classical equation to determine $\tilde{H}$ as well to implement the adiabatic quantum simulation under $\tilde{H}$). $\tilde{H}(\tau)$ becomes increasingly non-local with time, and we introduce a geometric locality heuristic to truncate  terms in $\tilde{H}$, which we compare to the original inexact QITE procedure.  
 We finish with some observations on the cost and practical implementation of the algorithm.

\section{Formalism}

\subsection{General theory}

 Consider a lattice system described by a Hamiltonian $H$. 
We desire an adiabatic Hamiltonian $\tilde{H}(\tau)$ whose ground-state at every $\tau$ is given by Eq.~\eqref{eq:imag_time_evolved_psi}. Consider an infinitesimally imaginary time evolved state from  $\tau$ to $\tau+d\tau$:
\begin{equation}\label{eq:infinitesimal_psi_evolution}
    \ket{\Psi(\tau+d\tau)}     = \ket{\Psi(\tau)} - d\tau \; Q H \ket{\Psi(\tau)} + O(d\tau^2),
\end{equation}
where $Q=1-\ket{\Psi(\tau)}\bra{\Psi(\tau)}$ projects out the ground-subspace. Now, suppose $\ket{\Psi(\tau)}$ is the ground-state of $\tilde H(\tau)$, one should determine $\tilde H(\tau+d\tau) = \tilde{H}(\tau) + \delta \tilde H$ such that $\ket{\Psi(\tau+d\tau)}$ is its ground-state: perturbation theory determines the ground-state of $\tilde H(\tau+d\tau)$ as $\ket{\Psi(\tau)} + [ \tilde E_0 - \tilde H(\tau) ]^{-1} Q \ \delta \tilde H \ket{\Psi(\tau)}$, where $\tilde E_0$ is the smallest eigenvalue of $\tilde H$. We will be working with evolution schemes for $\tilde H$ that start with and maintain $\tilde E_0 = 0$ (more on this below).
Thus using \eqref{eq:infinitesimal_psi_evolution} we should have:
\begin{equation} \label{eq:delta_tilde_H}
    \delta \tilde H \ket{\Psi(\tau)} = d\tau \, \tilde H H \ket{\Psi(\tau)} + O(d\tau^2),
\end{equation}
where we have used $\tilde{H} Q = Q \tilde H =\tilde H$.

Eq.~\eqref{eq:delta_tilde_H} is the main equation the adiabatic Hamiltonian $\tilde{H}$ should satisfy. However, it does not uniquely determine $\delta \tilde H$, and so there are many generating equations for $\tilde{H}$.
A simple choice is 
\begin{equation}\label{eq:naive_time_deriv}
\frac{d\tilde{H}}{d\tau} = \tilde H H + H \tilde H,    
\end{equation}
where we have used $\tilde H \ket{\Psi(\tau)} = 0$ {(see above Eq.~\eqref{eq:E_tilde_solution} for justification)} and added the term $H\tilde H$ to make the right-hand side Hermitian. This has the formal solution $\tilde{H}(\tau) = e^{H \tau} \tilde{H}(0) e^{H \tau}$ (where $\tilde{H}(0)$ is chosen so that $\Psi(0)$ is its ground-state with zero eigenvalue). 
The above scheme can, in principle, be implemented  as a hybrid quantum-classical algorithm, where $\tilde{H}$ is first determined by the classical integration of Eq.~\eqref{eq:naive_time_deriv}, and then used to implement {adiabatic} state evolution quantumly. (In other words, we carry out Hamiltonian dynamics with $\tilde{H}(\tau)$ along the adiabatic path $\tau = 0 \ldots \beta$, which, if performed slowly (see below) will guarantee that one remains in the ground-state of $\tilde{H}(\tau)$; the dynamics can be translated into a circuit using standard Hamiltonian simulation techniques~\cite{bauer2020quantum}). As is clear, this procedure does not involve any feedback from the quantum simulation, and thus does not involve tomography, unlike the original QITE. 

 {In addition, the above procedure can be used to generate an adiabatic Hamiltonian $\tilde{H}(\tau)$ to prepare states beyond $e^{-\beta H}|\Psi(0)\rangle$, generalising to the class of states $\mathcal{T} e^{- \int_0^\beta d\tau' L(\tau')}|\Psi(0)\rangle$, where $L(\tau')$ is an arbitrary operator that need not even be Hermitian. This can be seen by considering each time-step of propagating $L(\tau)$ separately and decomposing $L$ into a sum of Hermitian and anti-Hermitian parts $L_{\text{H}} + L_\text{AH}$. The time evolution reads $d\tilde{H}/d\tau = \{ \tilde{H}(\tau) ,L_\text{H}(\tau) \} + [ \tilde{H}(\tau) , L_\text{AH}(\tau)]$. Thus a wide class of states, beyond ground-states or physical imaginary time-evolved states, become accessible through an adiabatic state evolution (assuming a non-vanishing gap).}

However, this naive scheme has some potential problems. One set is analogous to that encountered in the original quantum imaginary time evolution scheme, namely, the time evolution of $\tilde{H}$ renders it nonlocal (both geometrically and in terms of the lengths of the Pauli strings in $\tilde{H}$) and increasingly complicated. For example, even if $H = \sum_\alpha h_\alpha$ and $\tilde H = \sum_\beta \tilde h_\beta$ are geometrically local, the time derivative introduces geometrically nonlocal terms like $h_\alpha \tilde{h}_\beta$, and the number of such terms grows exponentially with time (the exponent in the rate of increase of Pauli strings with significant coefficients depends on the details of $h_\alpha$) up till the maximum number of $4^L$ where $L$ is the size of the lattice.
This renders both the classical determination of $\tilde{H}(\tau)$, and the quantum implementation of state evolution under $\tilde{H}(\tau)$ inefficient for large $\tau$.

There is also a second set of problems
arising from the norm of $\tilde{H}$ and its spectrum. We use the symbol $|\tilde{\phi}_i(\tau)\rangle$ to denote the instantaneous eigenstate of $\tilde{H}(\tau)$ with eigenvalue $\tilde{E}_i(\tau)$ (if $\tilde{H}$ implements the imaginary time evolution perfectly, then $|\tilde{\phi}_0(\tau)\rangle = |\Psi(\tau)\rangle$).
Taking expectation values of Eq.~\eqref{eq:naive_time_deriv} with $|\tilde{\phi}_i \rangle$,
the eigenvalues of $\tilde H$ evolve as $\frac{d \tilde E_i}{d\tau} = 2 \tilde E_i \langle{\tilde{\phi}_i} |H| {\tilde{\phi}_i} \rangle$. If $\tilde H$ is initialized with zero ground-state energy, the evolution will keep it vanishing.
However, the other eigenvalues evolve as
\begin{equation}\label{eq:E_tilde_solution}
\begin{aligned}
\tilde E_i (\tau) &= \exp\left[2 \int_0^\tau d\tau'  \langle{\tilde{\phi}_i (\tau')} |H| {\tilde{\phi}_i(\tau')} \rangle \right] \tilde E_i(\tau=0)
\end{aligned}
\end{equation}
To see the potential problems with this, consider the example where $H$ has a finite spectrum with all eigenvalues above (or below) $0$. In that case, we clearly see that the eigenvalues $\tilde{E}_i, i>0$ are always growing (shrinking) with time, potentially exponentially fast. Thus there is the possibility for numerical issues at long times in determining $\tilde{H}$ and implementing evolution under it, as we now discuss.

The integration of the classical differential equation for $\tilde{H}$
and the corresponding time evolution under $\tilde{H}$ 
will carry some finite numerical error which depends on $||\tilde{H}||$. 
 This means that rather than 
obtaining the exact $\tilde{H}(\tau)$, we obtain $\tilde{H}(\tau) + V$; if one uses e.g. a finite-order Runge-Kutta method, then $||V|| \propto ||\tilde{H}||$. Depending on the dynamics of the eigenvalues, this
may introduce a large deviation from the instantaneous ground-state, for example, if $|\langle \tilde{\phi_0} |V|\tilde{\phi_i}\rangle/\tilde{\Delta}_i| \gg 1$ (where $\tilde{\Delta}_i$ is the instantaneous gap to the $i$th state of $\tilde{H}$). 
Similarly, the quantum  adiabatic evolution time depends on $||\tilde{H}(\tau)||$: 
since $\max_s ||d\tilde{H}/ds/\tilde{\Delta}(s)^2|| \propto \beta \|H\| \, \max_\tau ||\tilde{H}(\tau)/\tilde{\Delta}(\tau)^2||$, the total adiabatic evolution time $T \propto \beta \|H\| \, \max_\tau ||\tilde{H}(\tau)/\tilde{\Delta}(\tau)^2||$,  which can diverge if the numerator is exponentially growing or the denominator is exponentially decreasing. 
As a result, we will be focused on studying the behavior of $\|\tilde{H}(\tau)\|$ and $\tilde{\Delta}(\tau)$ numerically below (especially their behavior with total imaginary time and length).
Finally, the implementation of Hamiltonian simulation under $\tilde{H}(\tau)$ also introduces errors {that grow polynomially with $||\tilde{H}||$} (see e.g.~Corollary 2 of \cite{childs2021theory}).

The appearance of such problems might be considered natural given that reproducing the imaginary time trajectory as $\beta \to \infty$ is equivalent to the (formally) intractable problem of exact ground-state preparation. 
 However, as 
faithful imaginary time evolution produces an exponentially decaying infidelity with the final state, the long-time algorithm may not be needed if sufficient fidelity is already reached. 
For finite imaginary time, as needed in thermal state simulation or to prepare the more general class of states discussed above, the  concerns are also less relevant. In addition, various heuristics may be used to ameliorate these above concerns. 
We now turn to the discussion of heuristics, before studying the potential issues at finite and at long imaginary times, through numerical simulations.


\begin{figure}[t]
 	\centering
    \begin{overpic}[height=0.375\textwidth]{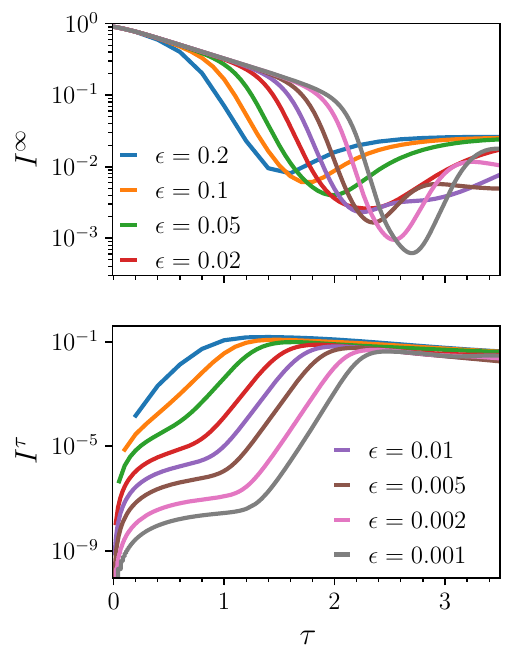}
        \put(0,175){(a)}
        \put(0,90){(b)}
      \end{overpic}
      \begin{overpic}[height=0.375\textwidth]{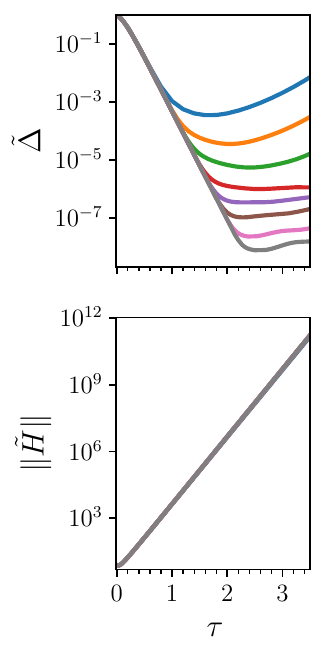}
        \put(0,175){(c)}
        \put(0,90){(d)}
      \end{overpic}
\caption{Adiabatic evolution under the Hamiltonian $\tilde{H}$ for a Heisenberg XXZ system of length $L=8$. $\tilde{H}$ is generated by integrating Eq.~\eqref{eq:naive_time_deriv} with second-order Runge-Kutta and time-step $\epsilon$, which is varied here.  $I^\infty$, $I^\tau$: infidelities with the ground-state of $H$ and with the exact imaginary time evolved state; $\tilde{\Delta}$: instantaneous gap of $\tilde{H}$; $||\tilde{H}||$: norm of $\tilde{H}$. 
The same color is used in all of the plots for each value of $\epsilon$, even though we have shown each $\epsilon$ only once.
\label{fig:equation_for_h}}
\end{figure}

\begin{figure}[t]
 	\centering
      \begin{overpic}[height=0.41\textwidth]{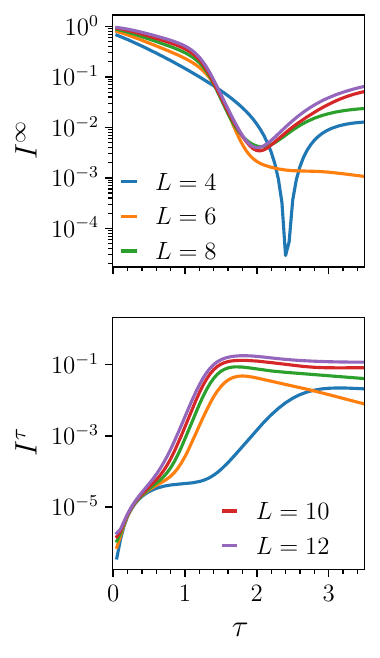}
        \put(5,195){(a)}
        \put(5,100){(b)}
      \end{overpic}
      \begin{overpic}[height=0.41\textwidth]{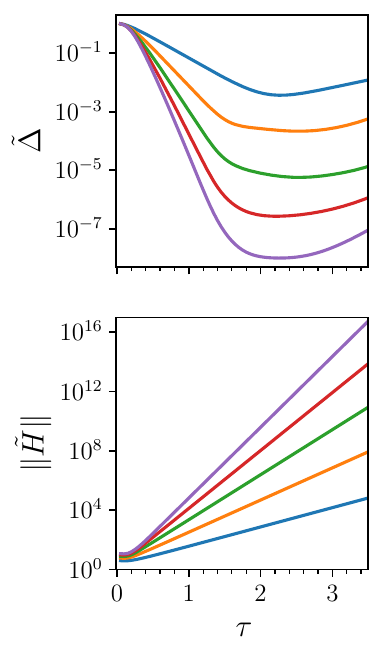}
        \put(5,195){(c)}
        \put(5,100){(d)}
      \end{overpic}
\caption{Adiabatic evolution under $\tilde{H}$ generated by Eq.~\eqref{eq:locality_preserving_deriv}, for the Heisenberg XXZ system as a function of chain length $L$. $\tilde{H}$ is generated using second-order Runge-Kutta and a timestep of $\epsilon = 0.05$. Quantities are same as in Fig.~\ref{fig:equation_for_h}. 
\label{fig:equation_for_h_L}}
\end{figure}

\subsection{Locality heuristic} 
To address the growing non-locality of $\tilde{H}$ we first write a modified generating equation for $\tilde H$, with separate differential equations for the individual terms $\tilde h_\beta$.
We choose $|\Psi(0)\rangle$ such that it is annihilated by each $\tilde h_\beta(0)$, and the evolution 
preserves this annihilation condition, analogous to Eq.~\eqref{eq:naive_time_deriv}.
We consider
\begin{equation}\label{eq:locality_preserving_deriv}
    \frac{d\tilde h_\beta }{d\tau}= \sum_\alpha \left\{ \tilde{h}_\beta , h_\alpha \right\}',
\end{equation}
where $\{a,b\}'$ denotes the anticommutator of $a$ and $b$ if they do not commute and zero if they do.
Eq.~\eqref{eq:locality_preserving_deriv} also satisfies
the adiabatic trajectory of Eq.~\eqref{eq:delta_tilde_H} because \eqref{eq:naive_time_deriv} and \eqref{eq:locality_preserving_deriv} only differ by terms that annihilate $\ket{\Psi(\tau)}$; so far no heuristic has been introduced.
The expression $\{ \tilde{h}_\beta , h_\alpha \}'$ means that $\tilde{H}$ no longer contains geometrically non-local terms: each 
$\tilde h_\beta$ grows its support from the contribution of terms $h_\alpha$ that overlap with the boundary of its support at every step. 

We can then introduce a heuristic to control the width of support. In particular, 
we can truncate summation over $\alpha$ in Eq.~\eqref{eq:locality_preserving_deriv} so that only a subset of terms $h_\alpha$ are retained for each $\tilde{h}_\beta$. More precisely, a neighborhood block is assigned to every $\tilde h_\beta$ term which is a region with a given spatial extent $w$ that surrounds the location of $\tilde h_\beta$ at $\tau = 0$; every $h_\alpha$ term that lies in the neighborhood block of $\tilde h_\beta$ is retained in Eq.~\eqref{eq:locality_preserving_deriv}. In this approximation, $\tilde{H}$ remains strictly {$w$-geometrically local} with time. We note that this above heuristic is different from the locality approximation in the original QITE algorithm, as it is a direct restriction on the operator, rather than the correlation length of the imaginary-time-evolved state. The relationship between the two is studied numerically below.

\subsection{Gauge degree of freedom}

We can introduce other modifications to the generating equation of $\tilde{H}$ which do not modify the imaginary time evolution trajectory but which can, in principle, affect $||\tilde{H}||$ and its spectrum. Consider, for example,
\begin{align}
        \frac{d\tilde h_\beta }{d\tau}= \sum_\alpha \left\{ \tilde{h}_\beta , h_\alpha \right\}' + f(\tilde{h}_\beta),
        \label{eq:modified_h_dynamics}
\end{align}
where $f(\tilde{h}_\beta(\tau))$ annihilates $\Psi(\tau)$.
This ensures that the ground-state of $\tilde{H}$ is a zero eigenstate for all $\tau$ ({although it does not ensure that it is always the ground-state}). We can then optimize $f(\tilde{h}_\beta)$ to control the gap and norm. 
Specifically, in tests below, we consider $f(\tilde{h}_\beta) = u_1 \tilde{h}_\beta - u_2\tilde{h}_\beta^2$. The $u_1$ term is equivalent to adding a constant shift to $H$ in Eq.~\eqref{eq:naive_time_deriv}.
While we do not expect this gauge choice to remove the fundamental difficulties of preparing states for infinite imaginary times, we can expect to improve the preparation of states for finite imaginary time.
In practice, we do not know how to choose $f(u)$ ahead of time, but it may be chosen in a heuristic manner, or $f(u$) may be chosen as part of a variational ansatz for $\Psi(\tau)$ at finite $\tau$.

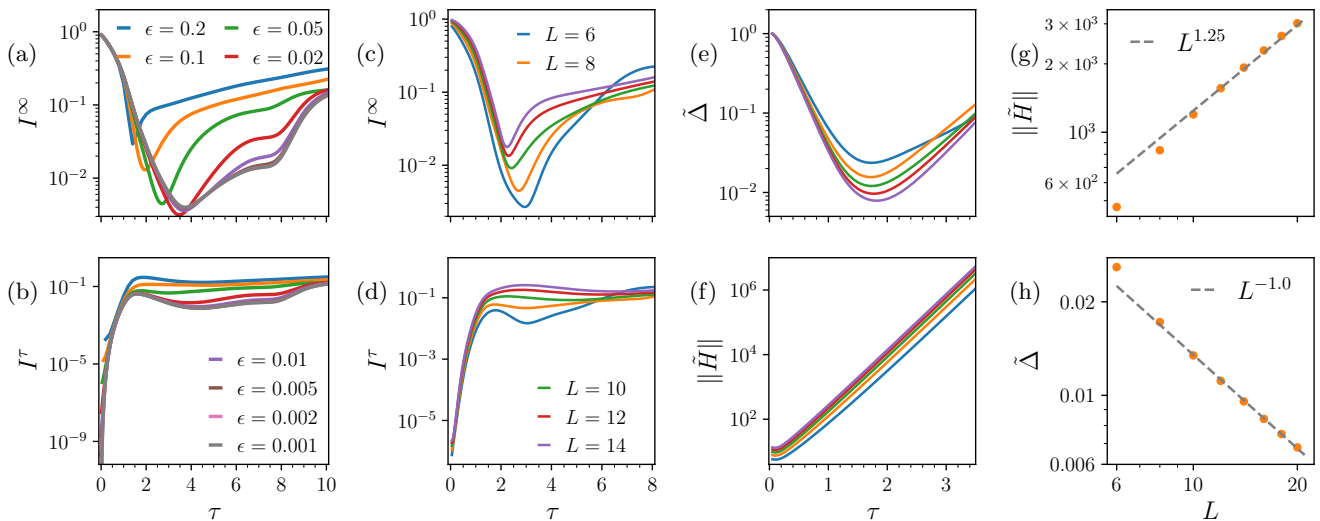
\begin{figure*}[!t]
 	\centering
    \begin{overpic}[height=0.4\textwidth]{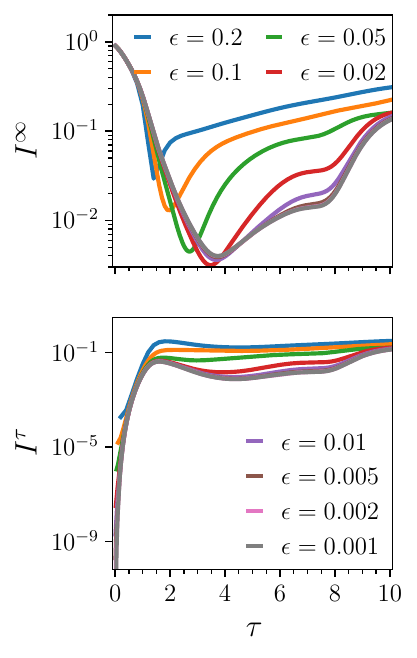}
        \put(0,180){(a)}
        \put(0,90){(b)}
      \end{overpic}
      \begin{overpic}[height=0.4\textwidth]{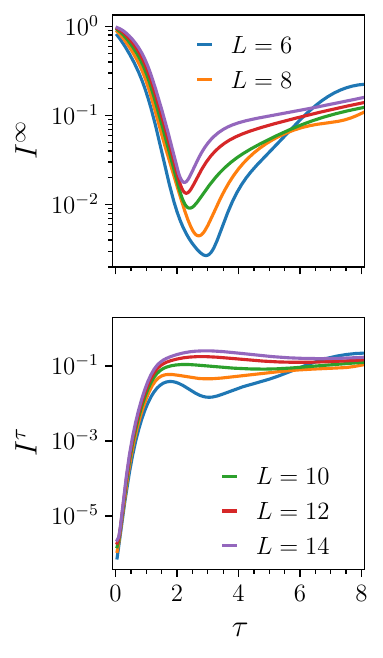}
        \put(0,180){(c)}
        \put(0,90){(d)}
      \end{overpic}
      \begin{overpic}[height=0.4\textwidth]{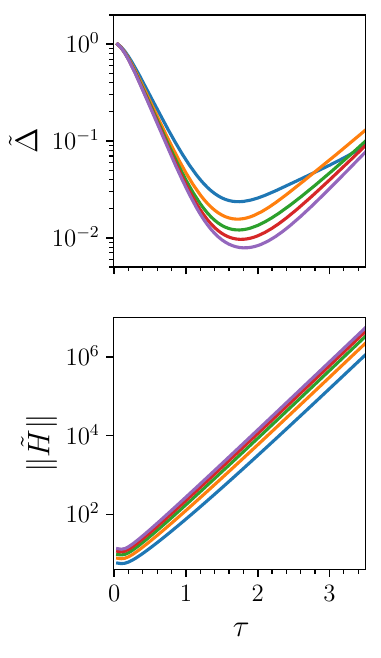}
        \put(5,180){(e)}
        \put(5,90){(f)}
      \end{overpic}
      \begin{overpic}[height=0.4\textwidth]{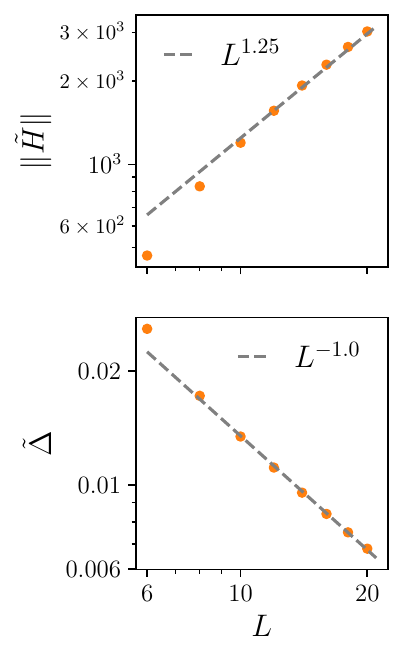}
        \put(5,180){(g)}
        \put(5,90){(h)}
      \end{overpic}
\caption{Adiabatic evolution under $\tilde{H}$ generated by Eq.~\eqref{eq:locality_preserving_deriv}, using a locality truncation with block size $w=5$, for Heisenberg XXZ chains of various lengths. 
(a) and (b) show infidelities for different values of $\epsilon$ and $L=8$. 
Note that the infidelities are defined in the same way as in Fig.~\ref{fig:equation_for_h}.
(c) and (d) show the behavior of the infidelities for different lengths, second-order Runge-Kutta, $\epsilon = 0.05$. (e) and (f) depict the gap and norm for different lengths, the same colors as the second column are used. 
(g) and (h) denote the norm and the gap of the adiabatic Hamiltonianas a function of length. The data is consistent with power-law behaviors for both.}
\label{fig:equation_for_h_w_5}
\end{figure*}

\section{Numerical Simulations} 

We now study the imaginary time trajectory generated by the $\tilde{H}$ dynamics, including the locality and gauge modifications described above. We classically propagate $\tilde{H}$ by a second-order Runge-Kutta method
\footnote{In the same fashion as Ref.~\cite{motta2020determining}, we take the effect of separate terms of $H$ into account separately for $\tilde H$, so each time step consists of a sequence of substeps, each one propagating for time $\epsilon$ with a single term of $H$.}
with a finite time step $\epsilon$ (error $O(\epsilon^3)$ per time-step; {note we use $\epsilon$ rather than $d\tau$ here to emphasize a finite step}) and then diagonalize $\tilde{H}(\tau)$ to study the trajectory of the ground-state $\Phi_0(\tau)$. We then  monitor various quantities, such as $||\tilde{H}(\tau)||$, the gap $\tilde{\Delta}(\tau)$, the infidelity with the
exact imaginary time propagated state $I^\tau(\tau) = 1- |\langle \tilde{\phi}_0(\tau)|\Psi(\tau)\rangle|^2$, and the infidelity with the ground-state of $H$, $I^\infty(\tau) = 1- |\langle \tilde{\phi}_0(\tau)|\Psi(\infty)\rangle|^2$,  {i.e. the outcome of the infinite imaginary time evolution.}

We study the antiferromagnetic Heisenberg XXZ model in one dimension with open boundary conditions:
\begin{equation}\label{eq:heisenberg_model}
    H_{\lambda_z} = \sum_j S^x_j S^x_{j+1} + S^y_j S^y_{j+1} + \lambda_z S^z_j S^z_{j+1},
\end{equation}
with $\lambda_z=2$ which results in an Ising-like anisotropy.  {To generate a simple initial state,}
we initialize the adiabatic Hamiltonian $\tilde H$ as a staggered transverse field:
$\tilde{H}(\tau=0) = \sum_j \left[(-1)^j S^x_j + \frac12 \right]$ (where each term in brackets 
separately annihilates the ground-state);  {the resulting initial state is thus a product state}. We then determine $\tilde{H}$ under the dynamics generated by Eqs.~\eqref{eq:naive_time_deriv},~\eqref{eq:locality_preserving_deriv}, and ~\eqref{eq:modified_h_dynamics}, and study various properties of $\tilde{H}$ and its instantaneous ground-state $\tilde{\phi}_0(\tau)$  {which is intended to produce the imaginary-time trajectory}. $\tilde{H}$ has the symmetry $\prod_i S^x_i$, under which $\tilde{\phi}_0(\tau)$ has a definite eigenvalue. Note that $H_{\lambda_z=2}$ has symmetry broken ground-states with long range order \cite{orbach1958linear,mikeska2004one}, but we take $\ket{\Psi(\infty)}$ to belong to the same symmetry sector as $\tilde{\phi}_0(\tau)$.

{In Fig.~\ref{fig:equation_for_h}} we first consider $\tilde{H}$ generated by Eq.~\eqref{eq:naive_time_deriv}, and the infidelities $I(\tau)$, $I^\infty(\tau)$, $||\tilde{H}(\tau)||$ and $\tilde{\Delta}(\tau)$ as a function of time-step $\epsilon$ used in the classical integration of Eq.~\eqref{eq:naive_time_deriv}. As seen, for all step-sizes, at early times the infidelity with the  ground-state of $H$, $I^\infty(\tau)$, decreases exponentially quickly, to $10^{-2}$ or less.  {(Although this is a small system, we note that achieving finite infidelities of $O(10^{-1})$ is already important for ground-state preparation applications as it enables variants of quantum phase estimation adapted to the near-term setting~\cite{ding2023even}).} 
 {The achieved infidelities decrease as $\epsilon$ is decreased towards 
an infinitesimal step size in the classical integrator, which would give $I^\tau(\tau)=0$ at all times and $I^\infty(\tau) \to 0$ at long times.} 
{For a finite step size $\epsilon$, we see}
$I^\infty(\tau)$ reaches a minimum value while $I^\tau(\tau)$ first increases with time, reaches a maximum and then plateaus. 
The time for the minimum $I^\infty$ is close to (slightly after) the time for the maximum $I^\tau$; we refer to this (approximate) time as $\tau_c$. 
$||\tilde{H}||$ increases exponentially, 
while $\tilde{\Delta}$ decreases exponentially until about $\tau_c$, before slowly increasing again. 
The above is consistent with our analysis that {for a finite time-step}
it is only possible to determine the adiabatic Hamiltonian accurately up to a maximum time $\tau_c$ .

\begin{figure}[!b]
    \centering
    \includegraphics[height=0.3\textwidth]{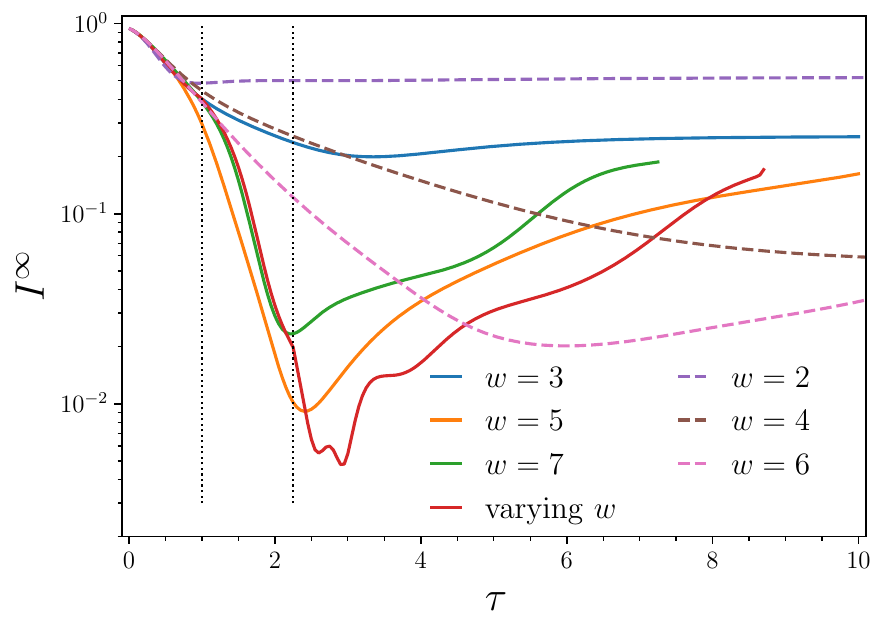}
    \caption{The infidelity with the desired ground-state for both the adiabatic QITE and the original QITE algorithm. Even values of $w$ correspond to the original QITE algorithm and the odd $w$ values and the varying case correspond to the adiabatic QITE. The time steps are taken as 0.05 for A-QITE and 0.08 for QITE. The varying $w$ case starts with $w=3$, then changes to $w=5$ and ultimately to $w=7$. The position of the transitions are shown by vertical dotted lines.}
    \label{fig:i_min_and_w}
\end{figure}

We next study  $\tilde{H}$ generated by  {the exact} Eq.~\eqref{eq:locality_preserving_deriv},  {i.e. without enforcing locality,} as a function of system size $L$,
and a fixed classical time step  $\epsilon=0.05$ in Fig.~\ref{fig:equation_for_h_L}. The results are similar to those using $\tilde{H}$;  {both formally generate the exact imaginary-time trajectory for infinitesimal step size.}
{Using a finite time-step, examining $I^\infty, I^\tau$}, we see the same behaviour of reaching a minimum $I^\infty$ and a maximum $I^\tau$ at some approximate time $\tau_c$.  
Surprisingly, this does not seem to change the maximum reliable propagation time, $\tau_c \approx 1.8$, between $L=6 - 12$.

\begin{figure*}[!!t]
 	\centering
    \begin{overpic}[height=0.41\textwidth]{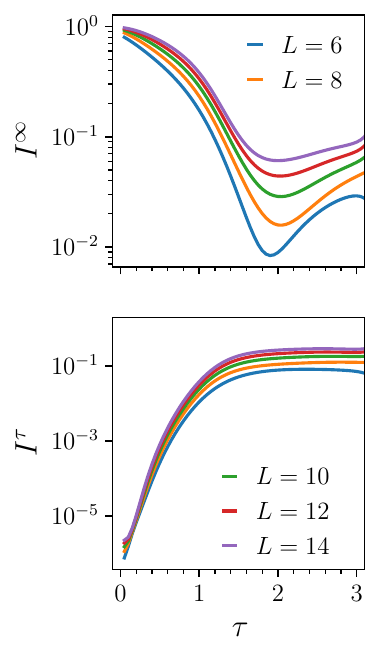}
        \put(5,195){(a)}
        \put(5,100){(b)}
      \end{overpic}
      \qquad
      \begin{overpic}[height=0.41\textwidth]{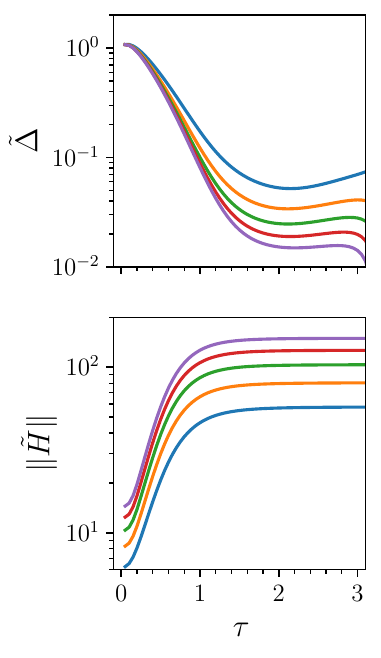}
        \put(5,195){(c)}
        \put(5,100){(d)}
      \end{overpic}
      \qquad
      \begin{overpic}[height=0.41\textwidth]{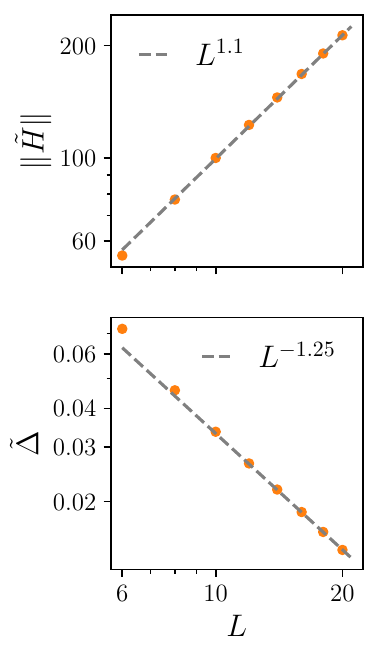}
        \put(0,195){(e)}
        \put(0,100){(f)}
      \end{overpic}
\caption{Adiabatic evolution under $\tilde{H}$ generated by Eq.~\eqref{eq:modified_h_dynamics}  with $u_1 = 2, u_2 = 0.5$, block size $w=5$, second-order Runge-Kutta, $\epsilon = 0.05$, for the Heisenberg XXZ system for various lengths. Quantities same as in Fig.~\ref{fig:equation_for_h}. (e) and (f) denote the norm and the gap of the adiabatic Hamiltonian as a function of length. The data is consistent with power-law behaviors for both.
\label{fig:equation_for_h_uful}}
\end{figure*}
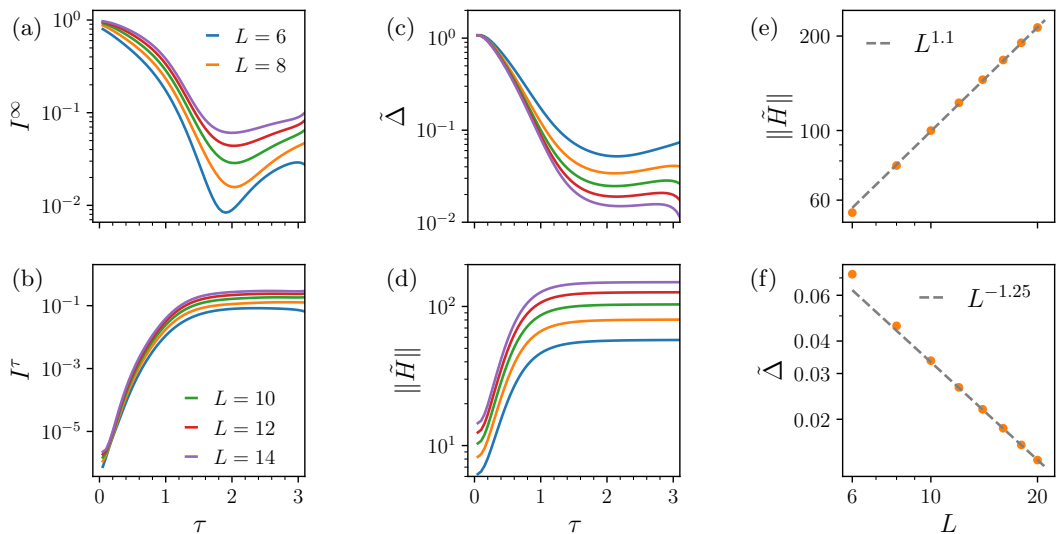

 {The above two numerical studies illustrate the formal exactness of the A-QITE procedure, as well as the numerical challenges in its implementation arising from the behaviour of $||\tilde{H}(\tau)||$ and $\tilde{\Delta}(\tau)$. We now study the impact of heuristics at controlling these quantities. We first}
apply the locality heuristic for the generation of $\tilde{H}$, implemented using Eq.~\eqref{eq:locality_preserving_deriv} with finite width $w=5$ in Fig.~\ref{fig:equation_for_h_w_5}. We study range of system sizes and integration time-steps.  {Although the finite width means that we no longer follow the imaginary-time trajectory exactly (even for infinitesimal time step), we still see exponential decay of the infidelity with the ground-state to useful values of $\sim 10^{-2}$ or less.}
{Overall, the  {infidelity} dynamics shows similar behaviour to previous examples, and at longer times there is a $\tau_c$} corresponding to a change in the behaviour of the infidelities and gaps. However, 
as the time-step goes to $0$, the minimum $I_\infty$ does not keep decreasing, because of the finite width error. 
 {From the perspective of the classical integration and quantum implementation, the dynamics generated under the locality heuristic is much more favourable than under the exact A-QITE $\tilde{H}$ dynamics. In particular, $\|\tilde{H}\|$ grows orders of magnitude more slowly with $\tau$, and (for fixed $\tau$) close to linearly with $L$, while $\tilde{\Delta}$ appears to decrease like $\mathrm{poly}(1/L)$.}

In Fig.~\ref{fig:i_min_and_w}, we show  $\tilde{H}$ generated by Eq.~\eqref{eq:locality_preserving_deriv}, using the locality heuristic with widths $3, 5, 7$, and timestep $\epsilon=0.05$, {as well as a protocol where the width is increased at increasing times}. For comparison, we also show data from
the original quantum imaginary time evolution scheme with widths $2, 4, 6$ and for time step $\epsilon=0.08$.
As expected, the best achievable fidelity with the
exact state $I^\infty(\tau)$ increases with $w$, however, the best $I(\infty)$ in fact decreases moving from $w=5$ to $w=7$. 
Compared to the original QITE scheme, the achievable infidelity appears to be better (i.e. lower) using the adiabatic Hamiltonian $\tilde{H}$ with a similar locality definition. 
Further work is required to establish a more precise relationship between these approximations.

We additionally consider the modified $\tilde{H}$ generated in  a different gauge using Eq.~\eqref{eq:modified_h_dynamics}. As an illustrative example, we consider the case $u_1 = 1/2$, $u_2=2$, with the locality constraint $w=5$. In Fig.~\ref{fig:equation_for_h_uful}, we see that although the behaviour of the infidelities $I_\infty$ and $I^\tau$ are qualitatively similar to what we have seen previously (corresponding to {$u_1=u_2=0$}) with a similar $\tau_c$, the detailed form is different; for example, the infidelity has a flatter plateau region.  {In addition, $||H||$
grows even more slowly with $\tau$ (with a flattened region between $\sim 1<\tau<3$) while still being close to linear in $L$, while $\Delta$ (at $\tau_c$) again appears to have a $\mathrm{poly}(1/L)$ behaviour.} 
This indicates that a suitable choice of $f(u)$ can meaningfully modify the dynamics and the achievable infidelities {at finite times}.

 {We now briefly discuss the implications for implementations in the near-term setting.
For a given width $w$ and Trotter time-step, the gate cost of implementing an A-QITE time-step is the same as that of QITE,
namely it is the cost to implement a time-step  under a $w$-qubit unitary, 
and importantly the A-QITE time-step requires no measurements. However, one difference with the original QITE is the need to maintain adiabaticity during the A-QITE simulation, which governs the time of Hamiltonian simulation for the given $\tau$.
The (limited) numerical data above on the scaling of $\tilde{H}$ and $\tilde{\Delta}$ when using the heuristic modifications in the Ising problem suggests that (for given $\tau$) the total adiabatic evolution time is polynomial in system size (as $||H||, ||\tilde{H}||$ scale like $\mathrm{poly}(L)$ and $\tilde{\Delta} \sim \mathrm{poly}(1/L)$). The Hamiltonian simulation is still likely the most challenging aspect of the algorithm for near-term hardware. However, for modest widths, e.g. $w=3$, implementing the A-QITE protocol is similar to implementing ASP with geometric 3-local Hamiltonians, which has already been demonstrated on near-term hardware using error mitigation and circuit recompilation~\cite{tan2023realizing}. 
Alternatively, it may be beneficial to relax the adiabaticity requirement by employing A-QITE within a version of the quantum approximation optimization algorithm (QAOA) \cite{farhi2014quantum,blekos2024review}, where $\tilde{H}$ provides the form of the operators in the ansatz.}

\section{Conclusions}
We have described an adiabatic state preparation protocol that implements the imaginary time evolution trajectory {without any need for quantum tomography or ancilla resources}. This hybrid algorithm involves a classical time integration to generate the adiabatic Hamiltonian,  {but does not require any measurements on the quantum system}. {When implemented faithfully, the algorithm leads to an exponential decrease of the infidelity with the ground-state of a desired Hamiltonian with adiabatic time. 
However, the cost of evolving exactly to long imaginary times grows rapidly with imaginary time both in the classical and quantum parts of the protocol.}
The growth in cost as a function of imaginary time arises from several sources, including the nonlocality of the derived adiabatic Hamiltonian. 
 {We introduce a  heuristic to control this nonlocality by truncating terms in the adiabatic Hamiltonian.} Another source of growing cost at long imaginary time is related to the norm of the adiabatic Hamiltonian and its gap.  {We show that modifying the generating equation of the adiabatic Hamiltonian can be used to control these quantities at finite imaginary times. Both heuristics enable one to propagate for short times and to observe a large improvement in the approximate ground-state.}


 {In the near-term quantum hardware setting, one advantage of the current approach is that each time-step can be implemented without measurements. On the other hand, preserving the strict adiabatic formulation may lead to long Hamiltonian simulation times for some problems.}

 {More generally, the A-QITE procedure we have introduced extends the types of states that be generated using an adiabatic simulation protocol; the target state need not be expressed as the ground-state of a known Hamiltonian, but rather can be expressed implicitly through the differential evolution of a known ground-state.} 
In addition, with respect to standard ground-state adiabatic state preparation, A-QITE expands the set of possible adiabatic paths. 
This potentially allows for introduction of new quantum adiabatic routines consisting of composite adiabatic paths (with A-QITE as one of them) with applications such as the introduction of novel adiabatic catalysts \cite{albash2018adiabatic,farhi2002quantum}, as directions to be explored in the future.

\bigskip

\section{Acknowledgments}

{We thank Yu Tong, Alex Dalzell and Anthony Chen  for helpful discussions.\\
KH  and GKC  were supported by the US Department of Energy, Office of Science, Basic Energy Sciences, under grant no. DE-SC-0019374. 
GKC acknowledges support from the Simons Foundation.}

\bibliography{a_qite.bib}

\end{document}